# The hydrodynamic singular vortex on the sphere and the Dirac monopole

**S.G. Chefranov** [1], **I.I. Mokhov** [1], **A.G. Chefranov** [2]

[1] A.M. Obukhov Institute of Atmospheric Physics RAS, Moscow, Russia
   E - mail: schefranov@mail.ru
[2] Eastern-Mediterranean University, Famagusta, North Cyprus

## Abstract

An exact correspondence is established between mathematical description of the single "elementary vortex" (EV) velocity on the sphere (Zermelo, 1902; Bogomolov, 1977) and the Dirac magnetic monopole (DMM) vector potential (Dirac, 1931). Similar mathematical analogy with DMM was noted only for the vortices in the quantum fluid $^3$He- A (Blaha, 1976; Volovik, Mineev, 1976). Singular elementary EV on a sphere is usually considered using compensating vortex field, uniformly distributed over the sphere surface. It is necessary to meet the Gauss-Kelvin theorem with zero integral vorticity over the entire surface of sphere. However, since Bogomolov (1977), (see also (Dritshel, Boatto, 2015)) uniform vorticity impact on the fluid particles dynamics is neglected and only the stream function of isolated singular EV is taken into account. It is not true because the stationary equations of hydrodynamics for a thin spherical layer of ideal incompressible fluid do not allow existence of a solution in the form of an isolated EV but allow solutions in the form of the antipodal EV pairs. It is also suggested the possibility of DMM existence only in the form of the point magnetic dipoles, consisting of two DMM with different signs of the magnetic charge.

## 1. Introduction

The problem of the existence of the isolated Dirac magnetic monopole (DMM) is fundamental for the theoretical physics [1-5]. Up to now, DMM are not being found in spite of the absence of any known restrictions on the possibility of the isolated existence of such objects in the electromagnetic field theory. In the present work, we state mathematical equivalence of this problem with the problem of existence of a single isolated "elementary vortex" (EV) on a sphere being "an isolated point vortex with constant (non-zero) vorticity density over entire sphere surface" [6]. In [7], long before and contrary to [6], only the pairs of EV with zero total intensity (corresponding to zero total vorticity density over the entire sphere) were considered as elementary vortex object on the sphere. As noted in [7], "only point vortices (infinitely thin vortex tubes) were considered, and the rest fluid flow was assumed non-rotational". Absence of the distributed over the sphere uniform vorticity gives an opportunity of introduction of the velocity potential to describe the fluid flow outside of the localizations of the point vortices, that allowed "very closely relating this part of hydrodynamics with the complex variable and logarithmic potential theories" [7].

Nevertheless, significantly later in [8], independent from [6], there was obtained the same as in [6] representation of EV stream function for isolated singular vortex with addition of uniform compensating vorticity that nowadays is actively used for the analysis of the point vortices on the sphere dynamics [9-12]. Moreover, after [8] and up to now the stream function of this singular EV is considered alone, without any influence of that compensating uniform vortex field on the dynamics of fluid particles even in the cases when the sum of the circulations is not zero for the system of such N singular EVs.

Herein, we show that EV velocity field, corresponding to a single EV stream function, cannot be a solution to the stationary hydrodynamics equations of the ideal incompressible fluid in the extremely thin spherical layer, considered in [6-13] in relation with the point vortices dynamics on the static and rotating sphere. At the same time, the solution of this equations may be represented as a velocity field of EV's pair with zero total circulation (or vorticity) over the entire sphere when these two point-vortices are antipodal (located in the diameter conjugated points of the sphere surface having the same by value but opposite by sign circulation values).

On the base of the found mathematical similarity between EV and DMM, we make a conclusion about the possibility of existence DMM only in the form of the DMM pairs, representing a point magnetic dipole of a new type (with magnetic field distribution different from the field of extremely small ring current).

## 2. Dirac Magnetic Monopole

According to [1], the DMM introduction in the quantum theory leads to the conclusion on possibility of observed quantization of the electric charge value with necessity requiring introduction of the vector potential $A_\mu$ of the magnetic field $\vec{H}$:

$$\vec{H} = rot\vec{A} \ . \qquad (2.1)$$

However, from (2.1), it follows a requirement on the absence of the magnetic charges:

$$div\vec{H} = 0 \ . \qquad (2.2)$$

In that relation, in [1], it is stated that the equation (2.2) cannot hold for DMM [1]: "There shall be (at least) one point on any surface embracing the monopole, where equation (2.2) is violated and where respectively cannot be introduced the vector potential $\vec{A}$ meeting equation (2.1). The points where the equation (2.1) is violated, at any time instance, form a line either continuing from the monopole to infinity, or ending on a monopole of the same or opposite sign charge. We call such a line – "string".

For DMM having the charge $g$ and placed at the coordinate system origin, the magnetic field is (see (2.1) in [1]):

$$H_i = H_{0i} = g\frac{x_i}{r^3}, r = \sqrt{\vec{x}^2} \ . \qquad (2.3)$$

In spherical coordinates, $(r,\theta,\varphi)$, expression (2.3) yields only one non-zero radial component

$$H_r = H_{0r} = \frac{g}{r^2}, H_{0\theta} = H_{0\varphi} = 0 \ . \qquad (2.4)$$

For the divergence of field (2.4) we have:

$$div\vec{H}_0 = \frac{1}{r^2}\frac{d}{dr}(r^2 H_r) = -g\Delta(\frac{1}{r}) = 4\pi g \delta(\vec{x}); \vec{x}^2 = r^2 . \quad (2.5)$$

In [1] as one of the possible variants of the magnetic vector potential selection for (2.1), it is proposed the potential having only one non-zero tangential component:

$$A_\varphi = A_\varphi^+ = -\frac{g}{r}ctg\frac{\theta}{2}; A_r = A_\theta = 0 . \quad (2.6)$$

Such a potential defines a string extending from the coordinate system origin to infinity along the direction specified by the angle $\theta = 0$ (i.e. directed from the sphere center to its North pole – angle $\theta$ is the complement to the latitude and is measured from the direction on the North pole).
 Another variant of the DMM vector potential considered in [1], is a string directed from the coordinate system origin to the diameter conjugated point, defined by the angle $\theta = \pi$:

$$A_\varphi = A_\varphi^- = \frac{g}{r}tg\frac{\theta}{2}; A_r = A_\theta = 0 . \quad (2.7)$$

There is incompliance of (2.4) with respect to (2.6) (or (2.7)) on the DMM string. Representation (2.4), actually, is not exact for any angle $\theta$ and is defined only with accuracy up to a contributor proportional to the Dirac delta-function, $\delta(\theta)$, characterizing singularity of the magnetic field in the points of the string.
 Actually, the magnetic field value defined from (2.1) for the potential (2.6) has the following form different from (2.4) (see Appendix):

$$H_r^+ = \frac{1}{r\sin\theta}\frac{\partial(A_\varphi^+ \sin\theta)}{\partial\theta} = -\frac{4\pi g}{r^2}\lim_{\theta_1 \to 0; \varphi_1 \to 0}(\frac{1}{\sin\theta_1}\delta(\theta-\theta_1)\delta(\varphi-\varphi_1)) + \frac{g}{r^2} . \quad (2.8)$$

In (2.8), it is assumed that the operation of the limit finding is performed only after integration in the definition of the Dirac delta-function as a generalized function. Note that the magnetic field in (2.8), contrary to the representation (2.4), already has zero-valued integral over a sphere of radius $r$.
 Relationship (2.8) is proved in Appendix (see [14] in this connection) taking into account for the potential (2.6) the relationship $div\vec{A}^+ = 0$ from which yields possibility of representation $A_\varphi^+ = -\frac{1}{r}\frac{\partial \psi_M^+}{\partial\theta}, A_\theta^+ = \frac{1}{r\sin\theta}\frac{\partial \psi_M^+}{\partial\varphi}$ of the vector potential in (2.6) via the function

$$\psi_M^+ = -g\ln(\frac{1}{1-\cos\theta}) . \quad (2.9)$$

Actually, integral over entire sphere of radius $r$ for the magnetic field in (2.8) is equal to zero in accordance with the Gauss theorem:

$$I^+ = r^2 \int_0^{2\pi} d\varphi \int_0^\pi d\theta \sin\theta H_r^+ = 0 . \quad (2.10)$$

If considering in (2.7) the charge of the opposite sign with respect to that in (2.6) (i.e. replacing in (2.7), $g \to -g$), then for (2.7), it can be introduced a function similar to (2.9) of the following form:

$$\psi_M^- = -g\ln(1+\cos\theta) . \quad (2.11)$$

Taking this into account, from (2.1) and (2.7) (using the substitution in (2.7), $g \to -g$), we get for the magnetic field of a DMM located in the coordinate system origin and having the opposite sign charge (with respect to the charge creating the magnetic field (2.8)):

$$H_r^- = \frac{4\pi g}{r^2} \lim_{\theta_1 \to 0; \varphi_1 \to 0} (\frac{1}{\sin(\pi - \theta_1)} \delta(\pi - \theta - \theta_1)\delta(\varphi - \varphi_1)) - \frac{g}{r^2}; \vec{H}^- = rot\vec{A}^-(-g) . \qquad (2.12)$$

Integral over entire surface of the sphere of radius $r$ of (2.12), similar to (2.10), is zero:

$$I^- = r^2 \int_0^{2\pi} d\varphi \int_0^{\pi} d\theta \sin\theta H_r^- = 0 . \qquad (2.13)$$

It is clear that for the sum of the fields, (2.8) and (2.12),

$$H_r = H_r^+ + H_r^- = -\frac{4\pi g}{r^2} \lim_{\theta_1 \to 0; \varphi_1 \to 0} (\frac{1}{\sin\theta_1}\delta(\theta - \theta_1) - \frac{1}{\sin(\pi - \theta_1)}\delta(\pi - \theta - \theta_1))\delta(\varphi - \varphi_1) . \qquad (2.14)$$

For (2.14) the Gauss theorem also exactly holds since the integral over the sphere to (2.14) is zero.

Such a total magnetic field corresponds to the sum of the functions (2.9) and (2.11) in the form

$$\psi_M = \psi_M^+ + \psi_M^- = -g\ln(\frac{1+\cos\theta}{1-\cos\theta}) . \qquad (2.15)$$

From (2.15), it is defined also the vector potential of the total magnetic field of the point magnetic dipole being the sum of two DMM's of the opposite signs having the following form

$$A_\varphi = -\frac{1}{r}\frac{\partial \psi_M}{\partial \theta} = -\frac{2g}{r\sin\theta} . \qquad (2.16)$$

Divergence of the magnetic field of the point magnetic dipole according to (2.14) is

$$div\vec{H} = -8\pi^2 g\delta(\vec{x})\lim_{\theta_1 \to 0; \varphi_1 \to 0}(\frac{1}{\sin\theta_1}\delta(\theta - \theta_1) - \frac{1}{\sin(\pi - \theta_1)}\delta(\pi - \theta_1 - \theta))\delta(\varphi - \varphi_1) . \qquad (2.17)$$

From (2.17), it follows already also that the integral $\int d^3x div\vec{H} = 0$ contrary to the integral for the magnetic field divergence in (2.5). The distribution of magnetic field (2.12) is different from the field which is produced by the small ring current on a large distance from it.

### 3. Point Vortex on the Sphere

Let us show that it is necessary considering not EV [6] as elementary vortex object on sphere but a pair of antipodal EV's for which the stream functions with accuracy up to the constant factor coincide with (2.15), and velocity field, with (2.16).

For the single vortex on the sphere, the stream function has [6, 8] the form (2.9), if using in (2.9) $g = -\frac{\Gamma}{2\pi}$, where $\Gamma$ is the circulation value. Then for the velocity field (on the sphere of radius r and with zero radial velocity component), from (2.9), we get the following expressions for the tangential velocity components:

$$V_\varphi^+ = -\frac{1}{r}\frac{\partial}{\partial\theta}\psi_M^+ (g = -\frac{\Gamma}{2\pi}) = \frac{\Gamma}{2\pi r}ctg\frac{\theta}{2}; V_\theta^+ = \frac{1}{r\sin\theta}\frac{\partial}{\partial\varphi}\psi_M^+ = 0, \quad (3.1)$$

coinciding with the expression (2.6).

Velocity field potential, corresponding to (3.1), has the following form (since $V_\varphi^+ = \frac{1}{r\sin\theta}\frac{\partial\Phi^+}{\partial\varphi}; V_\theta^+ = \frac{1}{r}\frac{\partial\Phi^+}{\partial\theta}$):

$$\Phi^+ = \frac{\varphi\Gamma}{4\pi}\cos^2\frac{\theta}{2}. \quad (3.2)$$

It is clear that (3.2) does not correspond to zero velocity component $V_\theta^+$ as in (3.1). It means that the stream function and velocity field in (3.1), obtained from (2.9) and (2.6) under substitution $g \to -\frac{\Gamma}{2\pi}$, correspond to the vorticity distribution which is different from zero outside of the localization point of the point vortex $\theta = 0$.

In the result, for the stream function of type (2.9), it is not possible introducing of conjugated with it velocity potential for which contour lines (where the potential is constant) must be strictly orthogonal to the stream lines (where the stream function is constant) [7].

For example, the contour line corresponding to the constant potential value $\Phi^+ = \frac{\Gamma}{4}$ in (3.2) is described by the function (see Fig.1 and Fig.2):

$$\varphi = \frac{\pi}{\cos^2\frac{\theta}{2}}. \quad (3.3)$$

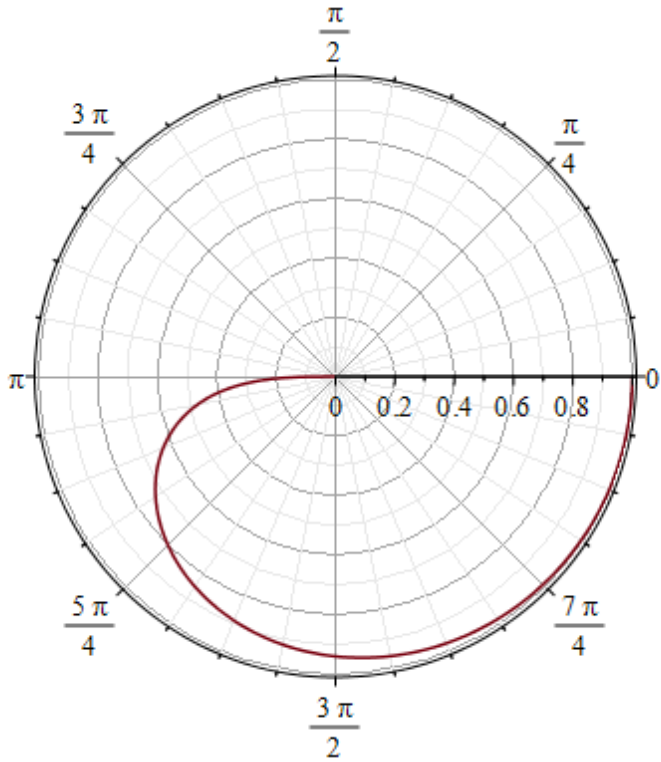

Fig. 1 Function (3.3) in Northern Hemisphere ( $0 \leq \theta \leq \pi/2$ )

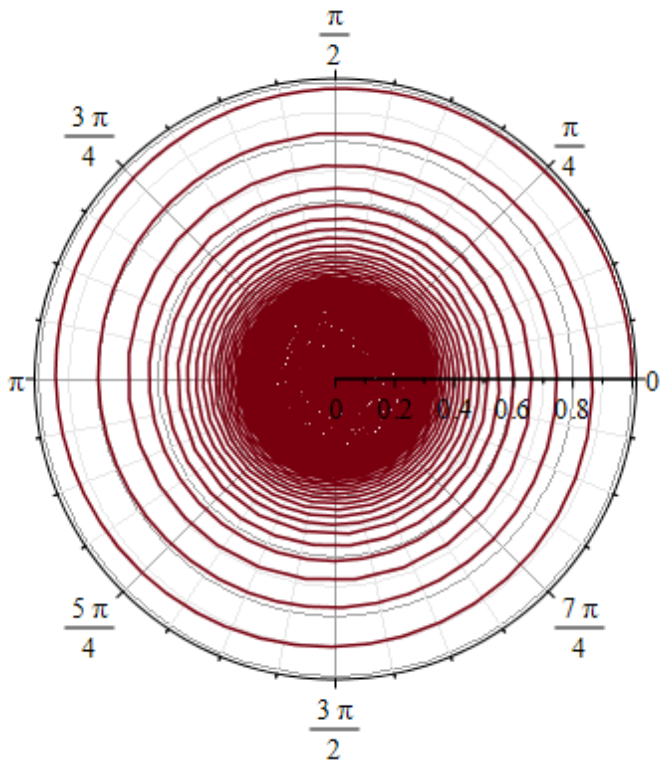

Fig.2 Function (3.3) in Southern Hemisphere ( $\pi/2 \leq \theta \leq \pi$ )

Thus, potential contour lines (3.2) are not, as noted above, orthogonal to the stream lines corresponding to the constant values of the function (2.9) at $\theta = const$.

On the other side, for the velocity field following from (2.15) and (2.16)

$$V_\varphi = -\frac{1}{r}\frac{\partial}{\partial\theta}\psi_M \, (g = -\frac{\Gamma}{2\pi}) = \frac{\Gamma}{r\pi\sin\theta} \,. \qquad (3.4)$$

For (3.4) we get the next expression for the potential:

$$\Phi = \Gamma\varphi \,. \qquad (3.5)$$

The dependence (3.5) already correctly describes the contour lines, $\Phi = const$, of the velocity potential which are actually orthogonal to the stream lines corresponding to the constant values of angle $\theta$ in (2.15).

Existence of the non-local vortex field distribution corresponding for the velocity field (3.1) (coinciding with (2.8) with accuracy up to a constant factor) is defined by existence of dependence of the velocity field circulation (on the latitude circle) from latitude. For the fixed value of the sphere radius, $r$, the value of the velocity field circulation (3.1) on the latitude circle defined by the angle value $\theta$ is

$$C^+ = r\sin\theta \int_0^{2\pi} d\varphi V_\varphi^+(\theta) = \Gamma\cos^2\frac{\theta}{2} \,. \qquad (3.6)$$

From (3.6), as noted in [6], for $\theta \to 0$, we get the circulation of the point vortex placed at $\theta = 0$ on the sphere. The velocity field circulation value of the point vortex placed at the point $\theta = 0$ does not depend on the angle $\theta$ only with accuracy $O(\theta^2)$ in the decomposition of the right-hand side of (3.6) to the power series of $\theta$.

In that limit, the stream function (2.9) and velocity field (2.6) correspond to the point vortex field on the plane for which the circulation value does not depend on the distance of the integration contour from the point vortex.

From the other side, the velocity field (3.4) circulation value is $C = 2\Gamma = const$ and does not depend on the angle $\theta$ now for any value of the angle.

In [13], it is proved that the velocity field (3.4) for a pair of antipodal point vortices exactly satisfies three-dimensional hydrodynamics equations for ideal incompressible fluid in the thin spherical fluid layer. At the same time, the velocity field (3.1) (corresponding to the stream function of type (2.9)) does not satisfy these equations.

Actually, stationary hydrodynamics equations for the thin spherical layer of the ideal incompressible fluid (with zero radial velocity field component), in particular, admit the following axial-symmetric representation in which only tangential component of the velocity field, $V_\varphi$, is non-zero [15]:

$$\frac{V_\varphi^2}{r} = \frac{1}{\rho_0}\frac{\partial p}{\partial r} \,, \qquad (3.7)$$

$$\frac{V_\varphi^2}{r}ctg\,\theta = \frac{1}{r\rho_0}\frac{\partial p}{\partial\theta} \,. \qquad (3.8)$$

The condition of compatibility of the equations (3.7) and (3.8), that is $\frac{\partial^2 p}{\partial r \partial \theta} = \frac{\partial^2 p}{\partial \theta \partial r}$, is:

$$\frac{1}{r}\frac{\partial V_\varphi}{\partial \theta} = \frac{\partial V_\varphi}{\partial r} ctg\,\theta \quad . \tag{3.9}$$

It is possible checking that the velocity field of a single EV (3.1) does not meet (3.9) and hence cannot correspond to any stationary solution of the hydrodynamics equation of the ideal incompressible fluid in the thin spherical layer.

At the same time, the velocity field (3.4), corresponding to the antipodal pair of point vortices on the sphere, already exactly satisfies (3.9) and hence is the solution of the hydrodynamics equations.

## 4. Analogy between Hydrodynamic and Electromagnetic Phenomena

Analogy between hydrodynamic and electric phenomena was noted by Helmholtz (1858) and by G. Maxwell who used assumption of hypothetical fluid dynamics when defining his theory [16]. Relatively recent works also use different variants of analogy between the Maxwell equations modifications and compressible medium hydrodynamics equations [17-19].

However, up to now, only one example is known of correspondence between the exact solution of the electrostatics problem for the point electric dipole potential and divergent-free velocity potential for the flow about the rigid sphere of the ideal incompressible fluid with the constant speed [20]. For magneto statics, also there is an example (in spite of it is not explicitly noted in [20]) of exact coincidence of the vector potential distribution created by the magnetic field of infinitely long solenoid in its cross plane and the point vortex field on the plane for the ideal incompressible fluid flow.

In the present work, for the first time, we state an exact mathematical correspondence between the velocity field (3.1) and magnetic field vector potential (2.6). In [17, 19], contrary to [18], the velocity field is also compared with the vector potential of the magnetic field when considering the analogy between hydrodynamic and electromagnetic field.

In that relation, we guess impossibility of existence of isolated DMM based on the obtained in the Section 3 proof of impossibility of existence of an isolated EV on the sphere surface.

At the first glance, the guess of impossibility of existence of isolated EV contradicts existing observation data on long-life isolated vortices in the atmosphere of Earth and other quickly rotating planets. However, from the other side, there are direct data [21] on existence of significant correlations of the very antipodal events related to the motion of the earth crust and also reanalysis data [22], pointing on the appearance of the long period (when averaging over the periods exceeding thirty years) correlations between the polar vortices of the Northern and Southern hemispheres. Conducted in [23] analysis that found existence of the cyclone-anticyclone vortex asymmetry (in relative frequencies of realization of sufficiently large-scale vortices of different signs) also does not exclude possibility of existence of the strong antipodal correlations. However, the existence of antipodal atmospheric vortex correlated with any same time long-living large-scale vortex is very difficult to detect due to the disturbances arising from different real stochastic factors. In [24], it is also proposed a "trick with the monopole" when while considering very long magnetic solenoid, the experimenter has access only to one end of the solenoid and he must find that the magnetic flux observed by him is fed by the solenoid. In [24], it is noted that if the experimenter has only classical charged particles for the determination of magnetic field then with their help he can find only the field exactly coinciding with the magnetic monopole field with charge $g$. Moreover, also when using quantum charged particles, he with the help of the Bohm-Aaronov effect (variant of the experiment with diffraction on two apertures) will not be able detecting solenoid if the following condition holds $eg = 0, \pm \frac{1}{2}, \pm 1,...$ , that coincides with the famous Dirac's quantization condition, where $e$ is the electron charge.

Note that up to now, any theoretical arguments in favor of absence of isolated DMM have not been defined. Importance of the guess obtained in the present paper can be illustrated by citation from [25]: "absence of monopoles can lead to conclusions significantly more important than those which might be demonstrated by their existence". In [25], first turn, arguments mentioned are related to symmetry of electric and magnetic phenomena which are used as a basis for theoretical conclusions on possibility of monopole existence.

**Conclusion**

A proof (see Section 3) is given of the absence of the hydrodynamics equations solution in the form of EV velocity field (3.1) introduced in [6, 8]. The EV velocity field (3.1) is similar to the DMM vector potential field (2.6). From the analogy between hydrodynamics and electrodynamics the impossibility of existence of single DMM (not united in antipodal dipole pairs) can be guessed. Direct contradiction with electromagnetic field equations is not found in spite of their analogy to the hydrodynamics equations.

The obtained results can be applied to the solution of one vortex problem in $^3He - A$ because of the same structure of superfluid velocity and EV velocity field. It may be verified by consideration of hydrodynamic equations stationary solutions for $^3He - A$ [26] (similar as we work herein with classical equations of hydrodynamics in Section 3).

**Appendix**

From (2.1) taking into account (2.9), one gets

$$\vec{H}^+ = rot\vec{A}^+ = (H_\theta^+ = 0; H_\varphi^+ = 0; H_r^+ = -\Delta \psi_M^+) , \quad (A.1)$$

$$-\Delta \psi_M^+ = -\frac{g}{r^2 \sin\theta} \frac{\partial}{\partial \theta} \sin\theta \frac{\partial}{\partial \theta} \ln(1-\cos\theta) . \quad (A.2)$$

From (A.1) and (A.2), the following expression follows $H_r^+ = -\frac{g}{r^2} \Delta_\theta (1-\cos\theta)$;

$$\Delta_\theta (1-\cos\theta) = \frac{1}{\sin\theta} \frac{\partial}{\partial \theta} \sin\theta \frac{\partial}{\partial \theta}(1-\cos\theta) = \frac{2}{\sin\frac{\theta}{2}\cos\frac{\theta}{2}} \frac{\partial}{\partial \theta} \sin\frac{\theta}{2}(1-\sin^2\frac{\theta}{2}) \frac{1}{\cos\frac{\theta}{2}} \frac{\partial \ln(\sin\frac{\theta}{2})}{\partial \theta}$$

(A.3)
Let us make in the right-hand side of (A.3) the following substitution

$$\rho = \sin\frac{\theta}{2} . \qquad (A.4)$$

Using (A.4), from (A.3), one gets

$$\Delta_\theta (1-\cos\theta) = \frac{1}{2}\left[\frac{1}{\rho}\frac{\partial}{\partial\rho}\rho(1-\rho^2)\frac{\partial\ln\rho}{\partial\rho}\right] = \frac{1}{2}\left[\frac{1}{\rho}\frac{\partial}{\partial\rho}\rho\frac{\partial\ln\rho}{\partial\rho}-2\right]. \quad (A.5)$$

Using the well-known relationship [14]

$$\Delta_{\vec{x}} \equiv \frac{1}{\rho}\frac{\partial}{\partial\rho}\rho\frac{\partial\ln\rho}{\partial\rho} = 2\pi\delta(\vec{x}); \vec{x}=(x_1,x_2) . \qquad (A.6)$$

$$x_1 = \rho\cos\varphi; x_2 = \rho\sin\varphi;$$

We use also the following formula

$$\delta(\vec{x}-\vec{x}_0) = \frac{1}{\rho_0}\delta(\rho-\rho_0)\delta(\varphi-\varphi_0) . \qquad (A.7)$$

Taking into account (A.4) and (A.7), and introducing the parameter, $\rho_0 = \sin\frac{\theta_0}{2}$, one gets from (A.6)

$$\Delta_{\vec{x}} = \lim_{\theta_0\to 0;\varphi_0\to 0}\frac{1}{\sin\frac{\theta_0}{2}}\delta(\sin\frac{\theta}{2}-\sin\frac{\theta_0}{2})\delta(\varphi-\varphi_0) . \qquad (A.8)$$

In (A.8), we use the following relationship

$$\delta(\sin\frac{\theta}{2}-\sin\frac{\theta_0}{2}) = \frac{2}{\left|\cos\frac{\theta_0}{2}\right|}\delta(\theta-\theta_0) . \qquad (A.9)$$

Using (A.9), we get from (A.3), (A.5), (A.6) and (A.8) the following expression for the radial magnetic field component

$$H_r^+ = -\frac{g}{r^2}\left[4\pi\lim_{\theta_0\to 0;\varphi_0\to 0}\frac{1}{\sin\theta_0}\delta(\theta-\theta_0)\delta(\varphi-\varphi_0)-1\right]. \qquad (A.10)$$

From (A.10), after opening the brackets, one gets (2.8).